\documentclass[conference]{IEEEtran}
\IEEEoverridecommandlockouts
\usepackage{cite}
\usepackage{amsmath,amssymb,amsfonts}
\usepackage{algorithmic}
\usepackage{graphicx}
\usepackage{textcomp}
\usepackage{xcolor}
\usepackage[a4paper, total={184mm,239mm}]{geometry}

\usepackage{tabularx}
\usepackage{booktabs,multirow}
\usepackage{array}
\usepackage{enumitem}
\usepackage[hidelinks,pdfauthor={Dinithi Jayasuriya, Devangshu Dutta, Aravind Saravanan, Nilesh Ahuja, Amanda Rios, Amit Trivedi},pdfcreator={},pdfsubject={},pdfkeywords={}]{hyperref}

\graphicspath{{figs/}}

\def\BibTeX{{\rm B\kern-.05em{\sc i\kern-.025em b}\kern-.08em
    T\kern-.1667em\lower.7ex\hbox{E}\kern-.125emX}}

\begin{document}

\title{Delta-Seeker: Closed-Loop Generative Search for Novel Circuit Design Discovery}

\author{
\IEEEauthorblockN{Dinithi Jayasuriya\IEEEauthorrefmark{1}, Devangshu Dutta\IEEEauthorrefmark{1}, Aravind Saravanan\IEEEauthorrefmark{1}, Nilesh Ahuja\IEEEauthorrefmark{2}, Amanda Rios\IEEEauthorrefmark{2}, Amit Trivedi\IEEEauthorrefmark{1}}
\IEEEauthorblockA{\IEEEauthorrefmark{1}University of Illinois Chicago, Chicago, IL, USA}
\IEEEauthorblockA{\IEEEauthorrefmark{2}Intel Corporation, USA}
}

\maketitle

\begin{abstract}
Generative models are increasingly used for circuit synthesis and optimization by proposing topologies, sizing devices, and exploring large design spaces. We investigate whether they can instead support \textit{design discovery}, where generated structures must depart from known designs while producing reproducible electrical behavior. We propose Delta-Seeker, a closed-loop circuit search framework that combines discrete diffusion for topology generation and mutation with structural checking, SPICE-in-the-loop sizing, Pareto selection, and feedback from prior search outcomes. The search controller uses progress and recurring structural failures to allocate evaluation effort and steer subsequent topology generation, while independent re-sizing and operating-corner analysis are performed only after search. In generator-only experiments, cross-generation feedback increases structurally valid children from 143 to 945 and reduces the faulty-candidate fraction from 0.925 to 0.761. In op-amp topology discovery at matched sizing cost, Delta-Seeker yields 54 designs that remain feasible in at least three of four independent re-sizings and 18 that additionally pass device-level checks and supply-temperature corner analysis, compared with 38 and 10 under fixed control. Removing cross-generation memory reduces independently robust discoveries from 54 to 43 and seven-corner robust discoveries from 40 to 27. In DC-DC converter search, nearest-parent analysis identifies a corpus-absent topology whose generated switching current path changes its nearest training-corpus parent from infeasible at all four operating corners to feasible at all four, while increasing median efficiency from 31.2\% to 98.96\%. These results show that closed-loop generative search can support circuit discovery beyond structural novelty by identifying new topologies with independently verified changes in circuit operation.
\end{abstract}


\section{Introduction}
\label{sec:intro}

\begin{figure}[t]
\centering
\includegraphics[
    width=\columnwidth
]{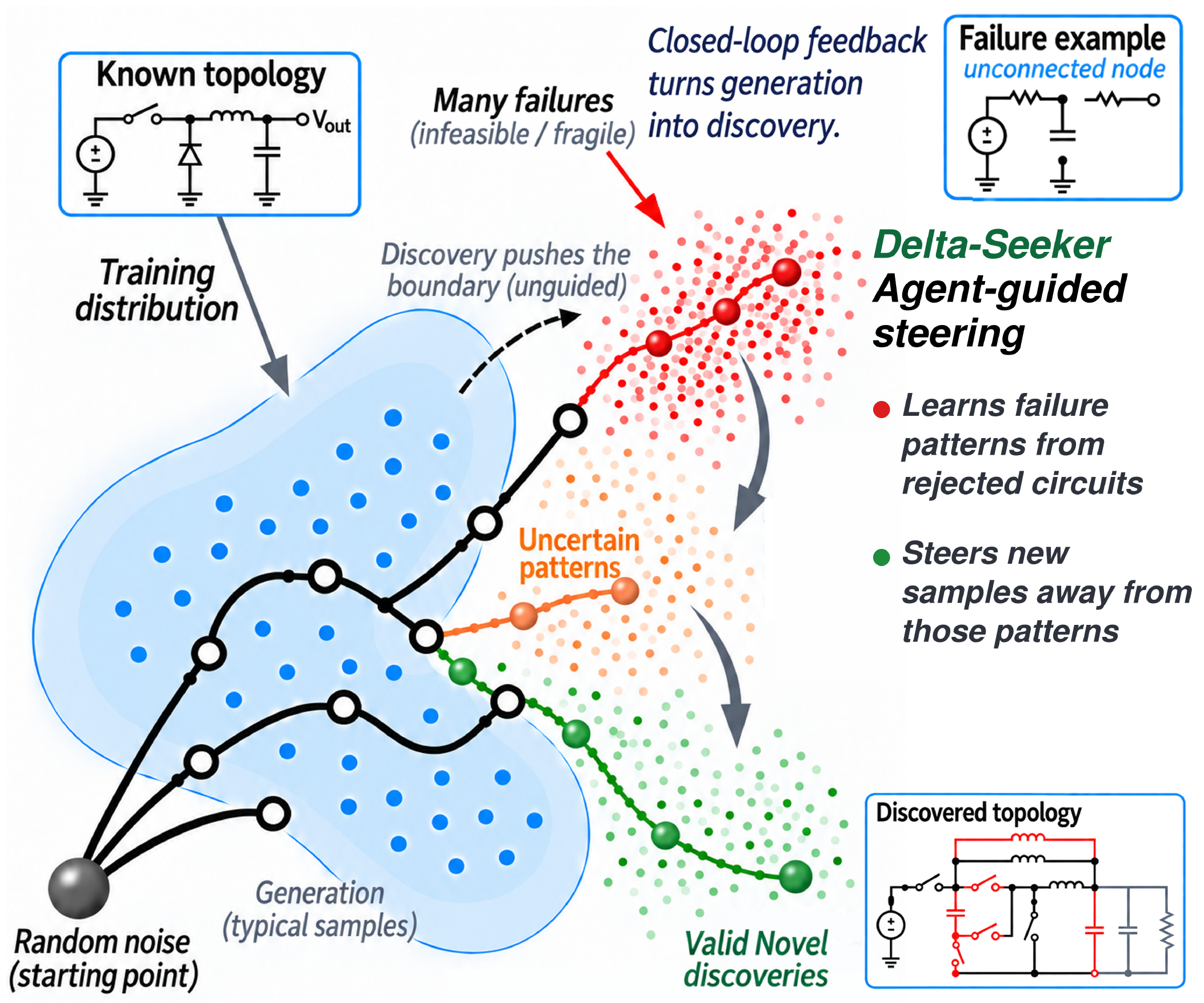}
\caption{\textbf{Delta-Seeker: From circuit generation to circuit discovery}. Generative sampling favors structures near the learned design distribution, while exploration farther from known designs increases the risk of invalid or poorly performing circuits. Delta-Seeker closes the loop by using outcomes from earlier candidates to steer topology generation toward distinct and useful designs.}\vspace{-10pt}
\label{fig:discovery_overview}
\end{figure}

Generative models are increasingly used to automate circuit synthesis and optimization by proposing topologies, sizing devices, and exploring large design spaces. Graph neural networks~\cite{dong2023cktgnn}, language models~\cite{chang2024lamagic,chang2025lamagic,ICLR2025_f0552f14,kim2026analogtobi}, reinforcement learning~\cite{vijayaraghavan2025autocircuitrl}, and diffusion models~\cite{liu2025diffckt,jayasuriya2026generation} have all been applied to analog and power-converter design. In this work, we investigate how generative design can move from \textit{design automation} toward \textit{design discovery}, where the objective is not only to generate feasible circuits, but to identify structures that depart from known designs while producing reproducible electrical behavior. This is challenging because departures from known topologies rapidly increase the risk of invalid connectivity, failed simulation, and poor performance.

We introduce Delta-Seeker (Fig.~\ref{fig:discovery_overview}), a closed-loop circuit discovery framework in which Delta denotes a verified departure from known designs in both circuit structure and operation. A diffusion model generates and mutates circuit topologies, while structural checks, SPICE-in-the-loop sizing, and Pareto selection evaluate feasibility, performance, and structural novelty. A state-dependent controller uses search progress and recurring structural failures to allocate additional evaluation effort and steer subsequent generations. Search outcomes therefore affect not only which candidates survive, but also which topologies are generated next. Unlike LLM-based EDA methods that directly generate or correct circuits~\cite{lai2025analogcoder,lai2026analogcoder}, select and size designs in multi-agent flows~\cite{shen2025atelier,habib2026aallm,wang2025multi}, reuse prior design knowledge~\cite{bao2026analogagent}, or operate design tools~\cite{wu2024chateda,wu2025divergent}, the controller in Delta-Seeker does not need to evaluate full performance.

\begin{figure}[t]
\centering
\includegraphics[
    width=\columnwidth
]{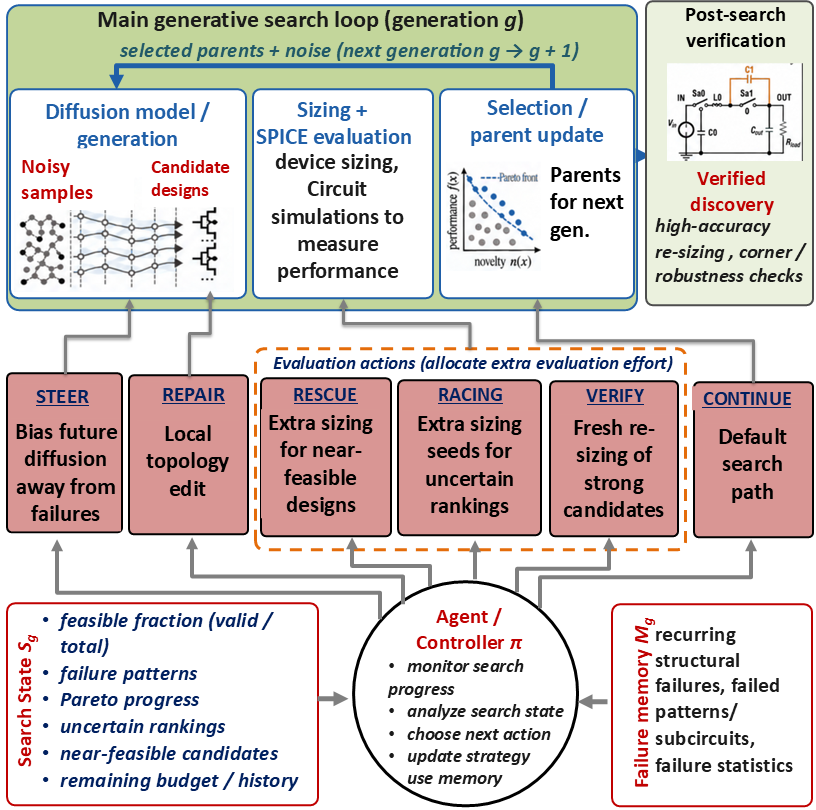}
\caption{Closed-loop search control in Delta-Seeker. From the current search state, remaining evaluation budget, and accumulated failure memory, the controller selects among topology steering, local repair, additional sizing or verification, and continuation. The selected action updates the search process, while steering modifies subsequent topology generation.}
\label{fig:closed_loop_control}\vspace{-10pt}
\end{figure}

We evaluate Delta-Seeker on op-amp and DC-DC converter topology discovery, which expose different failure modes. We choose these two classes because they span complementary discovery challenges, with op-amps stressing transistor-level feasibility and performance, and converters stressing whether structural novelty corresponds to a distinct and functional power-conversion topology. In op-amps, departures from known structures can quickly produce invalid connectivity or poor electrical performance. At comparable sizing cost, Delta-Seeker increases the number of topologies surviving independent re-sizing from 38 to 54 and the number additionally passing device-level checks and supply-temperature corners from 10 to 18. In converter discovery, structurally distinct graphs may instead be inactive, redundant, or functionally equivalent to known circuits. Nearest-parent and waveform analysis identify a corpus-absent topology whose generated switching current path changes its nearest training-corpus parent from infeasible at all four operating corners to feasible at all four, while increasing efficiency from 31.2\% to 98.96\%. Our contributions are:
\begin{itemize}[leftmargin=*, topsep=0pt, partopsep=0pt]

\item Delta-Seeker, a closed-loop generative circuit discovery framework that uses search outcomes to adapt both circuit evaluation and subsequent topology generation;

\item a cross-generation feedback mechanism that uses recurring structural failures and search progress to guide later exploration, with controlled ablations isolating its effect on topology generation and verified discovery yield; and

\item circuit-level validation across op-amp and DC-DC converter discovery using re-sizing, operating-corner analysis, circuit ablations, and equivalent-circuit reduction to distinguish structural novelty from verified changes in circuit operation.

\end{itemize}

\section{Closed-Loop Generative Circuit Discovery}
\label{sec:method}

In Fig. 2, Delta-Seeker operates as an iterative generate-evaluate-adapt loop. A diffusion model proposes circuit topologies, which are structurally checked, sized, and simulated in SPICE. The search records high-performing and near-feasible candidates together with recurring structural failures. This information allocates additional sizing or repair effort to current candidates and steers subsequent generations away from repeated failures. After search, selected designs are independently re-sized and verified to determine whether generated structural changes produce reproducible circuit behavior.

\vspace{4pt}
\noindent\textbf{A.~Topology Generation and Mutation.}
Each circuit is represented as a component-net graph whose terminals connect through net nodes and is converted into a SPICE netlist. Amplifiers are assembled from reusable motifs extracted from the training corpus, such as bias branches and transistor arrangements, while DC-DC converters are represented using switches, inductors, and capacitors. A discrete diffusion model $p_\theta$ is trained on valid circuit graphs to learn their structural distribution without simulated performance labels\cite{austin2021structured}. The first generation is obtained by denoising fully corrupted graphs. In later generations, selected circuits serve as parents and nearby variants are generated by partial corruption and denoising~\cite{meng2022sdedit}
\begin{equation}
K_{\gamma}(x' \mid x)
=
\int q(x_{\gamma T}\mid x)
\prod_{t=\gamma T}^{1}
p_\theta(x_{t-1}\mid x_t,c)\,
dx_{1:\gamma T},
\label{eq:kernel}
\end{equation}
where $q$ is the forward corruption process, $T$ is the diffusion length, and $\gamma$ controls mutation strength. Small $\gamma$ produces local topology changes, while $\gamma$ approaching one moves generation toward a new sample from the learned distribution. Structural constraints during denoising discourage invalid terminal connections and dangling nets. Delta-Seeker augments this generate-mutate process with cross-generation feedback so that failures observed earlier can alter later topology proposals.

\vspace{4pt}
\noindent\textbf{B.~Candidate Evaluation and Search.}
Each generated topology is first checked for structural validity and SPICE feasibility. Valid candidates are sized using an inner NSGA-II loop~\cite{deb2002fast}. The performance objective $f(x)$ is amplifier gain for op-amps and conversion efficiency for DC-DC converters, while a second objective measures structural novelty relative to the training set. For converters, using Weisfeiler-Lehman features~\cite{togninalli2019wasserstein} $\phi(x)$,
\begin{equation}
n(x) = 1 - \max_{y \in \mathcal{D}} \cos\!\big(\phi(x), \phi(y)\big),
\label{eq:novelty}
\end{equation}
where $\mathcal{D}$ contains the structurally distinct training circuits. The amplifier search uses the same graph-similarity principle. The outer loop performs Pareto selection over performance and structural novelty. Selected designs are mutated using~\eqref{eq:kernel} to form the next generation, while Pareto-optimal topologies are retained in the search archive.

\vspace{4pt}
\noindent\textbf{C.~Search Feedback and Adaptive Control.}
After each generation, Delta-Seeker summarizes feasible and near-feasible candidates, Pareto progress, uncertain rankings, recurring failures, previous search actions, and remaining evaluation budget. We denote the search state, action history, failure memory, and remaining budget by $S_g$, $H_g$, $M_g$, and $B_g$. The next operation is selected as
\begin{equation}
a_g = \pi\left(S_g,H_g,M_g,B_g\right).
\label{eq:policy}
\end{equation}
The selected operation uses the same structural-checking, sizing, and SPICE flow, and its outcome updates the search state before the next generation. The policy $\pi$ is implemented by an LLM that selects from a fixed set of search operations~\cite{yao2022react} but neither generates netlists nor evaluates circuit performance.

For existing candidates, \textsc{Rescue} allocates additional sizing effort to near-feasible designs, \textsc{Racing} evaluates additional sizing seeds when rankings are uncertain, \textsc{Verify} re-sizes strong candidates from fresh initializations, and \textsc{Repair} applies a bounded topology modification followed by checking, sizing, and simulation. \textsc{Continue} follows the default search path. \textsc{Steer} instead modifies subsequent topology generation. Failure patterns accumulated in $M_g$ bias later diffusion steps away from recurring faulty structures with strength $\lambda_g$, where $\lambda_g=0$ recovers the unbiased sampler. \textsc{Steer} may adjust $\lambda_g$, mutation strength $\gamma$ in~\eqref{eq:kernel}, and the number of children per parent. Changes to $\gamma$ and offspring count apply to the next generation, while $\lambda_g$ remains active until the next \textsc{Steer}. The diffusion model is not retrained, so failures observed in generation $g$ can directly alter the topologies proposed in generation $g+1$.

\vspace{4pt}
\noindent\textbf{D.~Verification.}
Search-time metrics alone do not establish discovery. After optimization, selected topologies are re-sized from fresh initial conditions and re-simulated independently of the search loop, and these results are not returned to the controller. For op-amps, verification includes repeated sizing convergence, device-body and bias connectivity, signal-current paths, gate-bias conditions, and supply-temperature operation with device sizes fixed. For DC-DC converters, selected topologies are evaluated across four operating corners and compared with their nearest training-corpus circuits. A topology is considered corpus-absent only when no equivalent graph is found after canonicalization, including checks after removal of one or two devices. Parent-child simulation, switching-waveform analysis, and equivalent-circuit reduction then determine whether the generated structural modification changes circuit operation. These post-search checks establish the Delta in Delta-Seeker as a structural departure from known designs accompanied by a reproducible change in circuit behavior.

\section{Op-Amp Topology Discovery}
\label{sec:amplifiers}

For op-amp topology discovery, Delta-Seeker seeks structures that differ from the training corpus while remaining electrically feasible after sizing and SPICE simulation. Larger departures from known amplifier structures increase invalid connectivity, failed operating points, and poor performance. We therefore measure whether search feedback increases distinct topologies that survive independent re-sizing and supply-temperature verification, rather than merely increasing search yield.

\vspace{4pt}
\noindent\textbf{A.~Experimental Setup.}
All methods use the same generator, SPICE evaluation and sizing flow, Pareto selection, and post-search verification. We compare them under a matched budget of approximately 3,000 sizing evaluations, allowing adaptive methods to allocate effort according to search state; realized cost ranges from 3,008 to 3,099 evaluations.

Base diffusion samples from the learned prior, while fixed control follows a predetermined search policy. Periodic steering uses the same failure memory and steering mechanism as Delta-Seeker but invokes \textsc{Steer} every five generations independent of search state (Fig.~\ref{fig:agent_trajectory}). Rule-adaptive control observes the same state, operations, memory, generator, and budget, but replaces the LLM policy with fixed if-then rules defined in advance. Delta-Seeker without memory retains state-dependent control but removes cross-generation failure memory, while the full method combines state-dependent control with memory-guided generation. All configurations use three seeds.

Post-search verification uses seven conditions: nominal;
$V_{\mathrm{DD}}\pm10\%$ at nominal temperature;
$-40^\circ$C and $125^\circ$C at nominal $V_{\mathrm{DD}}$;
and $V_{\mathrm{DD}}-10\%$ at $125^\circ$C and
$V_{\mathrm{DD}}+10\%$ at $-40^\circ$C.
Device sizes remain fixed across corners. We report both bias
re-trimming at each corner and the stricter case with nominal bias
also frozen.

\begin{table}[t]
\centering
\caption{Op-amp comparison under a common verification flow with independent fresh re-sizing of selected candidates.}
\label{tab:funnel}
\footnotesize
\setlength{\tabcolsep}{2.2pt}
\renewcommand{\arraystretch}{1.05}
\begin{tabular}{@{}lrrrrrr@{}}
\toprule
\textbf{Method} & \textbf{Sizing} & \textbf{Valid} & \textbf{Pareto} & \textbf{Fresh} & \textbf{$V,T$} & \textbf{Final} \\
 & \textbf{eval.} & \textbf{(\%)} & \textbf{archive} & \textbf{re-size} & & \\
\midrule
Base diffusion                & 3,027 & 22 & 14 & 8  & 5  & 3  \\
Fixed control                 & 3,074 & 92 & 59 & 38 & 31 & 10 \\
Periodic & 3,099 & 94 & 85 & 45 & 25 & 10 \\
Rule adaptive & 3,066 & 59 & 29 & 15 & 24 & 13\\
Delta-Seeker, no memory    & 3,008 & 92 & 81 & 43 & 27 & 12 \\
Delta-Seeker               & 3,024 & 95 & 99 & 54 & 40 & 18 \\
\bottomrule
\end{tabular}
\end{table}

\begin{figure}[t]
\centering
\includegraphics[width=\columnwidth]{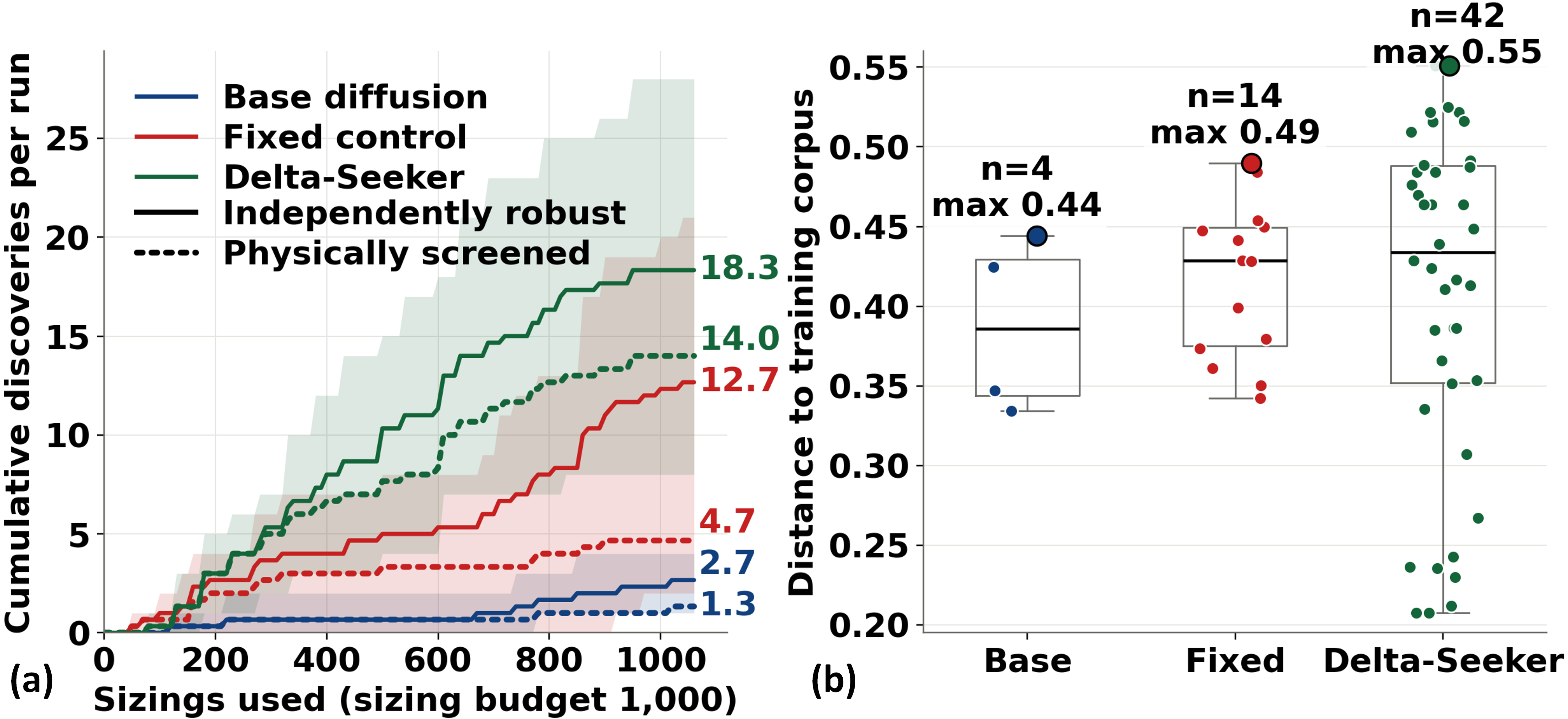}
\caption{Discovery yield under a fixed 1,000-sizing budget. \textbf{Left}, cumulative independently re-sized and device-checked designs as sizing evaluations are consumed. \textbf{Right}, structural distance of the device-checked designs from the training corpus.}\vspace{-10pt}
\label{fig:discovery_budget}
\end{figure}

\vspace{4pt}
\noindent\textbf{B.~Discovery Yield.}
Under the common evaluation flow in Table~\ref{tab:funnel},
Delta-Seeker achieves the highest verified discovery yield.
Of 99 archived topologies, 54 survive independent re-sizing and
18 also pass device and bias checks and all seven supply-temperature
conditions, versus 38 and 10 under fixed control. Periodic steering
and removing cross-generation memory both reduce this yield,
showing that state-dependent intervention and memory-guided
generation contribute to robust discovery. Structural novelty
remains comparable to fixed control, so the improvement does not
come from reduced novelty. As shown in Fig.~\ref{fig:discovery_budget}, a separate experiment with exactly 1,000 sizing evaluations confirms the trend under
a strictly matched budget, with Delta-Seeker yielding 14.0
device-checked designs per run versus 4.7 for fixed control and
1.3 for base diffusion. Median structural distance is 0.43 for both Delta-Seeker and fixed control, indicating that the higher yield results from more effective search rather than reduced novelty.

Across the ten verified designs in Table~\ref{tab:opamp_examples},
gain spans 27.4-37.9\,dB, GBW 2.6-885.3\,MHz, and phase margin
$53^\circ$-$96^\circ$, showing diversity in topology and operating
regime. All ten pass all seven supply-temperature conditions with
bias re-trimming, whereas only two to four remain feasible with both
device sizes and bias frozen. We therefore report these criteria
separately, since robustness of topology and sizing does not imply
robustness of a fixed bias point.

\begin{table}[t]
\centering
\caption{Representative verified op-amp discoveries. Fresh: feasible re-sizings; M: MOS devices; $+n$: passives; Nov.: WL distance; $\star$: Fig.~\ref{fig:verified_topologies}.}
\label{tab:opamp_examples}
\footnotesize
\setlength{\tabcolsep}{1.2pt}
\renewcommand{\arraystretch}{1.03}
\begin{tabular}{@{}lcccccccc@{}}
\toprule
\textbf{Design} &
\textbf{Dev.} &
\textbf{Gain} &
\textbf{GBW} &
\textbf{PM} &
\textbf{Power} &
\textbf{Fresh} &
\textbf{Corners} &
\textbf{Nov.} \\
& &
\textbf{dB} &
\textbf{MHz} &
\textbf{$^\circ$} &
\textbf{mW} &
& \textbf{trim/froz.} & \\
\midrule
D1$\star$  & 4M   & 32.1 & 120.9 & 92 & 3.36   & 4/4 & 7/7, 2/7 & 0.393 \\
D2$\star$  & 5M+4 & 30.5 & 62.2  & 96 & 9.80   & 3/4 & 7/7, 3/7 & 0.464 \\
D3$\star$         & 7M   & 37.9 & 885.3 & 53 & 31.49  & 3/4 & 7/7, 2/7 & 0.328 \\
D4         & 4M   & 34.0 & 257.5 & 89 & 42.27  & 3/4 & 7/7, 3/7 & 0.228 \\
D5$\star$          & 5M   & 33.8 & 2.6   & 91 & 14.71  & 3/4 & 7/7, 4/7 & 0.303 \\
D6         & 8M   & 31.6 & 481.7 & 88 & 287.20 & 3/4 & 7/7, 3/7 & 0.411 \\
D7         & 6M+2 & 29.4 & 84.9  & 73 & 5.02   & 3/4 & 7/7, 2/7 & 0.429 \\
D8         & 3M+1 & 29.3 & 189.2 & 94 & 11.30  & 4/4 & 7/7, 2/7 & 0.388 \\
D9         & 4M   & 27.8 & 108.8 & 96 & 6.84   & 3/4 & 7/7, 2/7 & 0.365 \\
D10        & 4M+1 & 27.4 & 207.8 & 88 & 14.43  & 3/4 & 7/7, 2/7 & 0.352 \\
\bottomrule
\end{tabular}
\end{table}

\begin{figure}[t]
\centering
\includegraphics[width=\columnwidth]{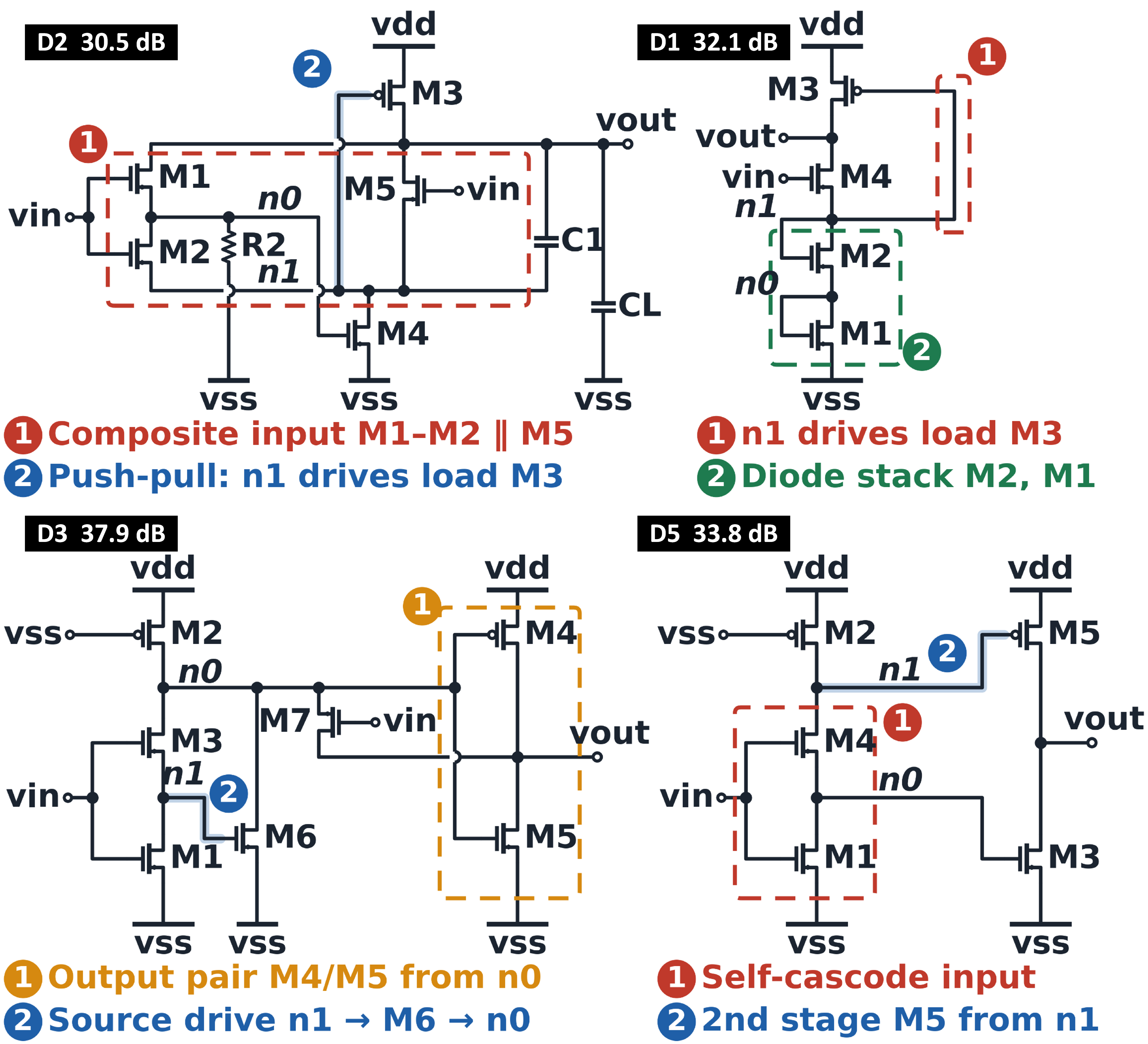}
\caption{Four representative Delta-Seeker op-amp discoveries after independent re-sizing: D1 uses push-pull load drive through the source-degeneration node; D2 combines a composite input branch, flipped-voltage-follower bias, and push-pull load drive; D3 is a two-stage amplifier with complementary output drive; and D5 combines a self-cascode input with second-stage drive. Colored annotations mark mechanisms examined by ablation.}
\vspace{-10pt}
\label{fig:verified_topologies}
\end{figure}

\vspace{4pt}
\noindent\textbf{C.~Representative Topologies and Circuit-Level Concepts.}
Table~\ref{tab:opamp_examples} summarizes ten verified discoveries spanning different gain, bandwidth, phase margin, power, and structural distance from the training set. Fig.~\ref{fig:verified_topologies} examines four representative topologies to determine what circuit mechanisms the generated connections realize and how they contribute to operation.

\vspace{4pt}
\begin{itemize}[leftmargin=*, topsep=0pt, partopsep=0pt, itemsep=3pt, parsep=0pt]

\item \textbf{\emph{D1, push-pull load drive through the source-degeneration node.}}
D1 uses four transistors with no passive components. Diode-connected M2 and M1 degenerate the source of input device M4, while the resulting node n1 also drives the PMOS load M3, coupling source degeneration directly to load drive. D1 is feasible in all four independent re-sizings. Holding the M3 gate at its DC value reduces gain by 0.6-3.7\,dB in every run, while moving it to the lower diode node leaves only two of four re-sizings feasible. Thus, the search identifies a distinct realization of push-pull load drive through the source-degeneration node.

\item \textbf{\emph{D2, composite input branch with push-pull load drive.}}
The M1-M2 branch forms a series-parallel composite transistor~\cite{galup1994series} in parallel with M5, while M4 and M2 form a flipped voltage follower~\cite{carvajal2005flipped}. Node n1 also drives PMOS load M3, producing push-pull operation~\cite{allen2012cmos}. Across three independent re-sizings, D2 reaches 23.0-30.5\,dB gain with up to $96^\circ$ phase margin. The discovery is therefore not a new primitive, but a generated combination of established circuit mechanisms absent from the nearest training examples.

\item \textbf{\emph{D3, two-stage amplifier with complementary output drive.}}
M3 and M1 form the first stage, while complementary devices M4/M5, both driven from n0, form a push-pull second stage. Holding either output-pair gate reduces gain by 3-10\,dB, confirming its role, while removing M1, M5, M6, or M7 and re-sizing reduces gain from 35.0 to 26.2-30.1\,dB. D3 reaches the highest verified gain among the four examples at 37.9\,dB, showing that the generated topology realizes a functional two-stage amplifier rather than a structurally novel but inactive graph.

\item \textbf{\emph{D5, self-cascode input with second-stage drive.}}
M4 and M1 form a self-cascode input structure~\cite{galup1994series}, and the input-stage output n1 drives PMOS M5 in the second stage. Holding the M1 gate reduces gain by 8-31\,dB, while holding M5 costs 5-12\,dB, confirming the input transconductance and second-stage paths. By contrast, holding the generated n0-to-M3 connection can increase gain, showing that not every generated connection contributes positively. D5 reaches 20.3-33.8\,dB and passes four of seven corners with bias frozen.

\end{itemize}

\begin{figure}[t]
\centering
\includegraphics[width=\columnwidth]{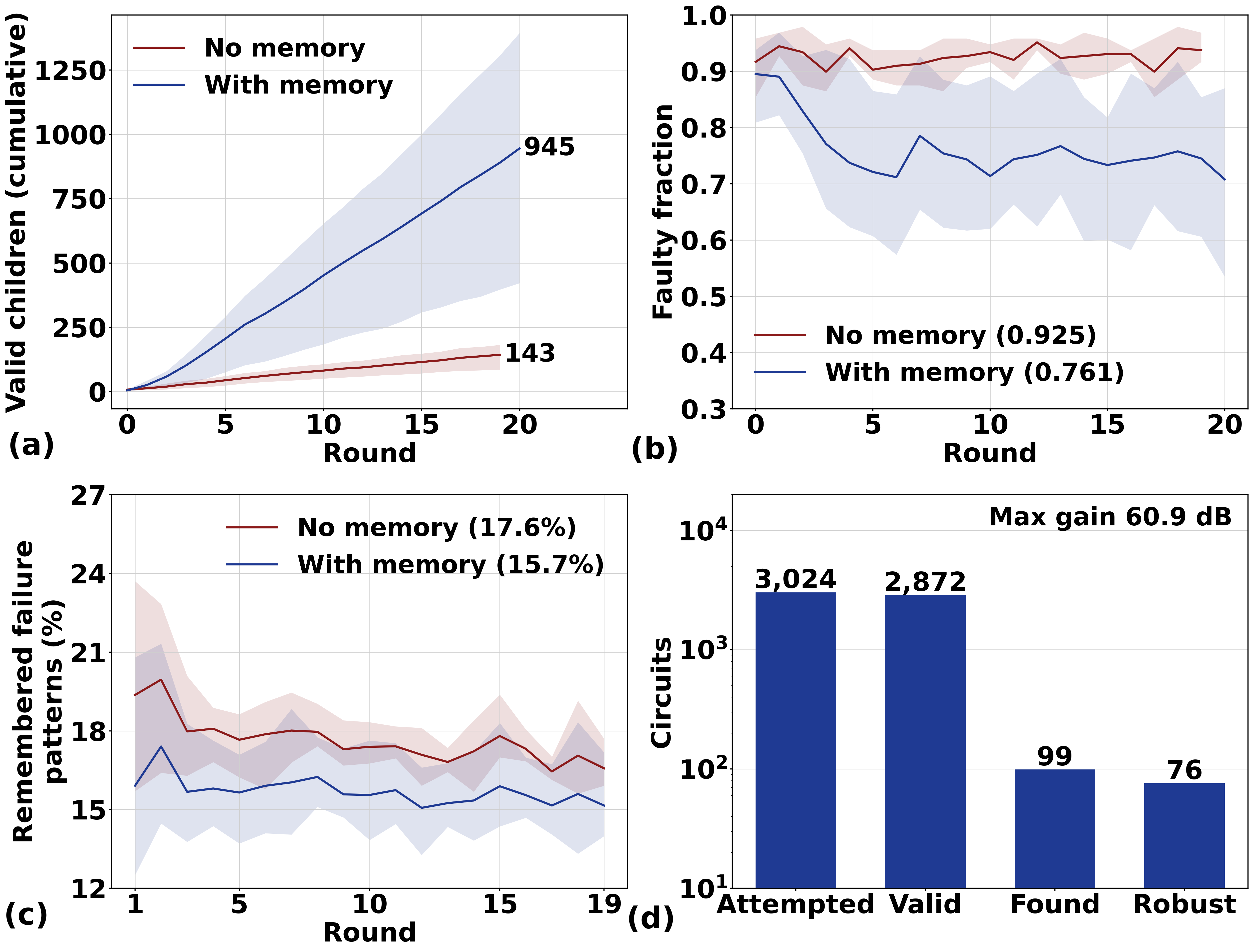}
\caption{Effect of cross-generation failure memory. Panels (a)-(c) isolate its effect on topology generation over 20 rounds and three seeds, without sizing or search control. Panel (d) shows the full Delta-Seeker search; 76 of 99 archived designs pass the end-of-run robustness re-score ($\geq 2/4$), while independent fresh re-sizing with the stricter $\geq 3/4$ criterion confirms 54 designs in Table~\ref{tab:funnel}.}\vspace{-5pt}
\label{fig:memory_steering}
\end{figure}

\begin{figure}[t]
\centering
\includegraphics[width=\columnwidth]{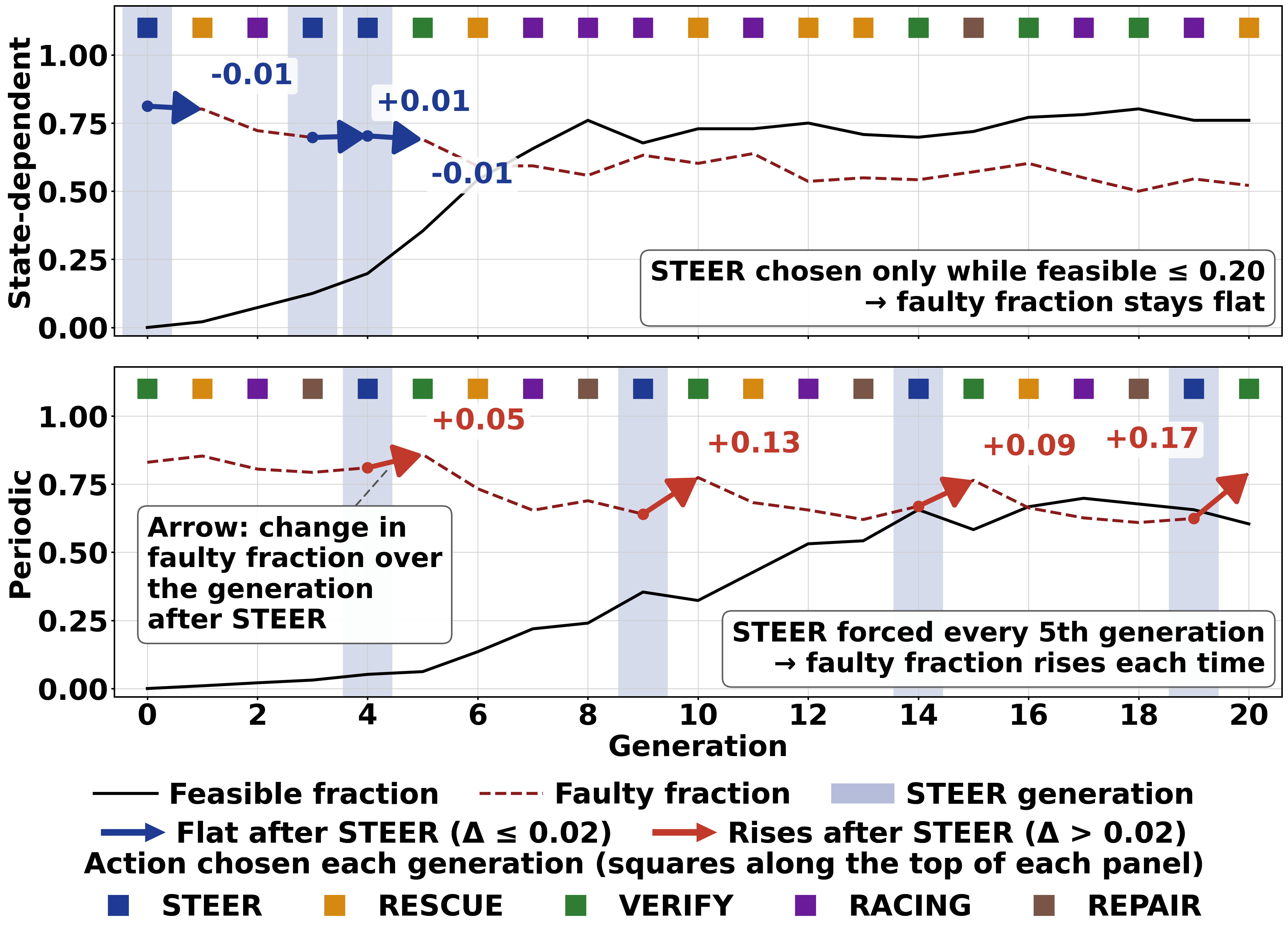}
\caption{Search trajectories under state-dependent control and periodic steering for the same seed. Solid and dashed curves show feasible and faulty-candidate fractions. Markers show operations, and shading indicates \textsc{Steer}.}\vspace{-5pt}
\label{fig:agent_trajectory}
\end{figure}

\vspace{4pt}
\noindent\textbf{D.~Effect of Cross-Generation Memory.}
As shown in Fig.~\ref{fig:memory_steering}, generator-only sampling
isolates cross-generation memory without sizing, Pareto selection,
or search control. Memory reduces the structural-fault fraction from
0.925 to 0.761, increases cumulative valid children from 143 to 945
per seed, and reduces recurring failure patterns from 17.6\% to
15.7\%. With no downstream selection, these results show that
failure history directly changes subsequent topology generation.

A full-search ablation removes cross-generation memory while
retaining adaptive evaluation. Restoring memory increases independently
re-sized designs from 43 to 54, seven-corner robust designs from 27 to
40, and the Pareto archive from 81 to 99. A physical screen of
device-body and bias connectivity, signal-current paths, and gate-bias
conditions passes 21 designs with memory and 23 without, indicating
that memory primarily improves robust downstream discovery rather than
intermediate verification.

\vspace{4pt}
\noindent\textbf{E.~Effect of State-Dependent Control.}
We separate the effect of \emph{when} feedback is applied from \emph{how} search actions are selected. Periodic control and Delta-Seeker use the same generator, failure memory, and operations, but periodic control intervenes on a fixed schedule while Delta-Seeker responds to the observed search state. As shown in Fig.~\ref{fig:agent_trajectory}, Delta-Seeker applies \textsc{Steer} when feasibility is low and later shifts effort toward \textsc{Rescue}, \textsc{Verify}, and \textsc{Racing}. Across the intervention study, state-dependent steering is followed by a mean change in faulty-candidate fraction of only $+0.009$, compared with $+0.096$ under periodic steering. At matched sizing cost, this translates to 54 independently re-sized and 18 final designs, versus 45 and 10 under periodic control, showing that feedback timing materially affects verified discovery yield.

We further compare against a rule-adaptive controller that observes the same state, operations, failure memory, generator, and budget, but selects actions using fixed if-then rules defined in advance. Under the matched budget, rule-adaptive control produces 13 final designs, compared with 18 for full Delta-Seeker. Thus, state-dependent intervention improves over periodic steering, while the LLM-based policy provides an additional gain over a hand-written state-dependent controller.

We also test sensitivity to the controller language model using the same three seeds and nearly identical sizing cost. GPT-5 yields more Pareto designs (106 vs.~99), independently re-sized designs (63 vs.~54), and seven-corner designs (42 vs.~40) than GPT-5-mini, but only 6 pass final verification compared with 18 for GPT-5-mini. Inspection shows that the rejected designs are dominated by two structural wiring defects: NMOS bodies tied to ground rather than $V_{\mathrm{SS}}$, and input nodes connected to non-gate terminals. Thus, higher intermediate yield does not necessarily translate to verified discovery, underscoring the need for independent post-search verification.

\vspace{5pt}
\noindent\textbf{F.~~Comparison Scope.}
Reported results from prior analog topology generators use different datasets, PDKs, objectives, and validity definitions~\cite{ICLR2025_f0552f14,kim2026analogtobi,lai2025analogcoder}. We therefore use them as methodological context rather than as numerical baselines. The discovery-yield claims above are based on the controlled configurations in Table~\ref{tab:funnel}, which share the same generator checkpoint, PDK, SPICE flow, sizing procedure, and verification criteria.

\section{DC-DC Converter Topology Discovery}
\label{sec:converters}

The converter study tests whether structural novelty changes circuit operation. With only three to six devices, corpus-absent graphs may be inactive, redundant, or functionally equivalent to known converters. We therefore use operating-corner, nearest-parent, waveform, and equivalent-circuit analyses to verify generated topologies.

\vspace{4pt}
\noindent\textbf{Converter Search Setup.}
The LaMAGIC corpus~\cite{chang2024lamagic} contains phase-a/b switches, inductors, and capacitors connected among $V_{\mathrm{IN}}$, $V_{\mathrm{OUT}}$, ground, and internal nets. After filtering invalid records and duplicate duty labels, it yields 37,514 distinct three-to-six-device graphs used to train the diffusion model conditioned on duty. At $D=0.5$, 71-83\% of random samples pass structural checks, but none of 100 meets the 2:1 conversion target and 20-30\% are pass-through structures, showing that structural validity does not ensure correct power conversion.

Candidates are simulated at $V_{\mathrm{IN}}=100$\,V and
100\,kHz with fixed component and parasitic models. Feasibility
requires $V_{\mathrm{OUT}}$ within 5\% of 50\,V, ripple below
5\%, efficiency $\geq90\%$, and a valid duty response across
nominal, heavy-load, light-load, and low-line conditions. Structural
checks reject floating or unbound connections, shoot-through paths,
shorted outputs, and missing input-to-output paths. Search uses a
population of 48 for 20 generations with efficiency and structural
novelty objectives. Four-corner-feasible designs are canonicalized
and compared with the training corpus after one- and two-device
removals. Corpus-absent designs are then compared with their nearest
training parents using waveform, line/load, tolerance, and
equivalent-circuit analyses to test whether generated elements alter
circuit operation.

\begin{figure}[t]
\centering
\includegraphics[width=1\columnwidth]{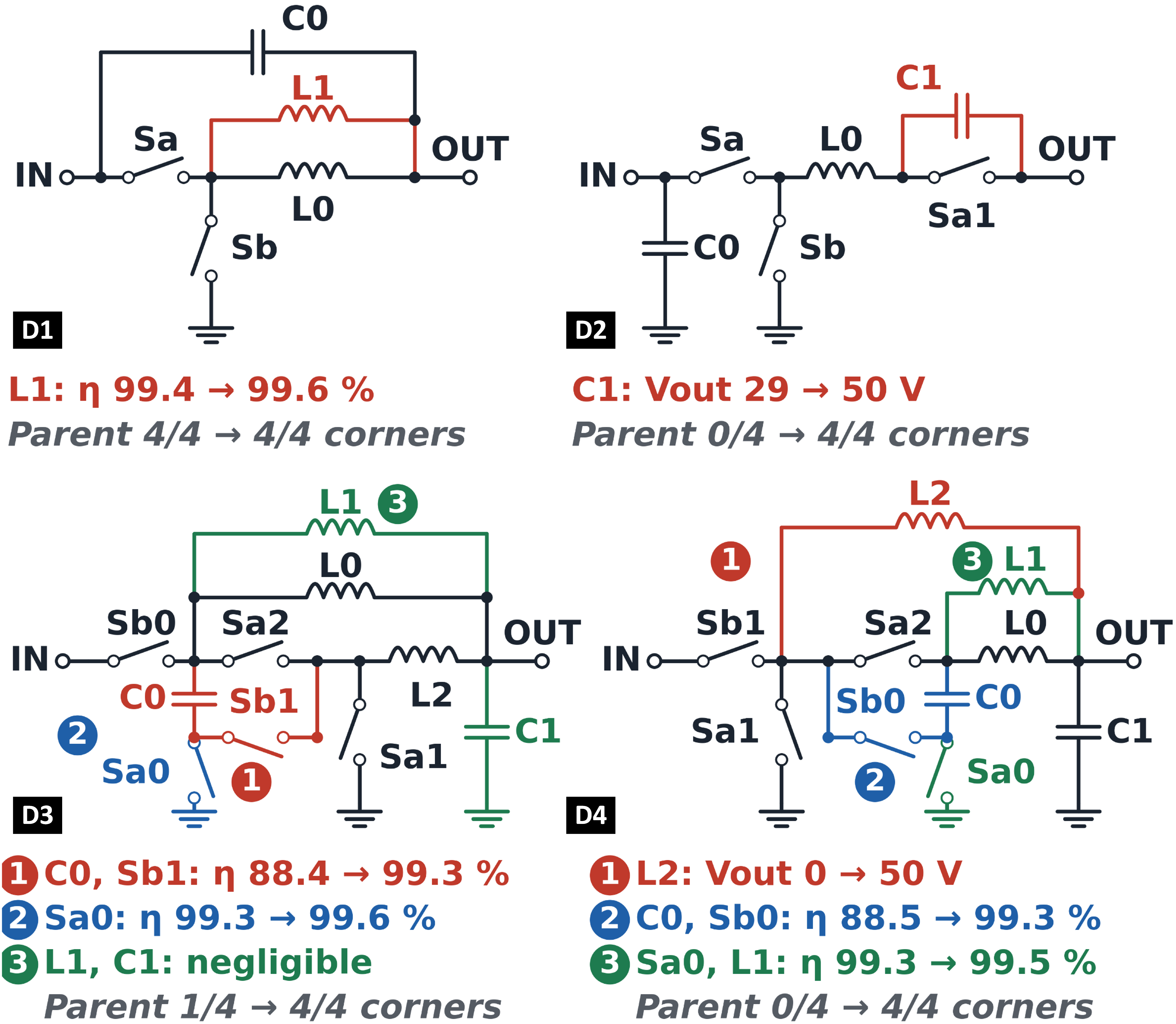}
\caption{Four corpus-absent converters and their nearest training-corpus parents.
Red indicates generated changes and grey the load. D1 improves a
functional parent, D2 adds a switching current path enabling four-corner
feasibility, and D3-D4 repair infeasible parents but reduce to known
buck-family structures.}
\label{fig:discovered_converters}
\end{figure}

\begin{table}[t]
\centering
\caption{Parent-child comparison for the four converters in Fig.~\ref{fig:discovered_converters}
(D1-D4: left to right, top to bottom) across four operating corners.}
\label{tab:conv_parent_child}
\scriptsize
\setlength{\tabcolsep}{1.25pt}
\renewcommand{\arraystretch}{1.08}
\begin{tabularx}{\columnwidth}{@{}c>{\raggedright\arraybackslash}Xcccc@{}}
\toprule
\textbf{Design} & \textbf{Generated change} & \textbf{Corners} & \textbf{Med. $\eta$} & \textbf{Worst $\eta$} & \textbf{Ripple} \\
& & & \textbf{(\%)} & \textbf{(\%)} & \textbf{(mV)} \\
\midrule
D1 & $L_1$ parallel to $L_0$ & $4/4\!\rightarrow\!4/4$ & $99.37\!\rightarrow\!99.60$ & $98.80\!\rightarrow\!99.28$ & $44\!\rightarrow\!88$ \\
D2 & $C_1$ across $S_{a1}$ & $0/4\!\rightarrow\!4/4$ & $31.18\!\rightarrow\!98.96$ & $16.82\!\rightarrow\!97.99$ & $251\!\rightarrow\!88$ \\
D3 & $C_0,S_{b1}$ link; $S_{a0},L_1,C_1$ & $1/4\!\rightarrow\!4/4$ & $88.43\!\rightarrow\!99.64$ & $79.76\!\rightarrow\!99.41$ & $147\!\rightarrow\!131$ \\
D4 & $L_2$; $C_0,S_{b0}$ link; $S_{a0},L_1$ & $0/4\!\rightarrow\!4/4$ & $0\!\rightarrow\!99.45$ & $0\!\rightarrow\!99.08$ & $--\!\rightarrow\!131$ \\
\midrule
Ref. & Textbook buck & $4/4$ & 99.37 & 98.80 & 87 \\
\bottomrule
\end{tabularx}
\end{table}

\noindent\textbf{Converter Discovery Yield.}
As shown in Table~\ref{tab:conv_yield}, under the matched converter budget, Delta-Seeker produces a similar
number of four-corner feasible designs to fixed control, but substantially
more structurally distinct and electrically active discoveries.
Across five seeds, training-absent designs increase from 72 to 100
and designs with active generated elements from 48 to 60.
Thus, the benefit becomes clearer after novelty and functional screening
rather than at feasibility alone.
\begin{table}[t]
\centering
\caption{Converter discovery across methods under a matched budget over five seeds.}
\label{tab:conv_yield}
\footnotesize
\setlength{\tabcolsep}{2.4pt}
\begin{tabular}{lccc}
\toprule
\textbf{Method} & \textbf{4-corner} & \textbf{Training-} & \textbf{Active} \\
 & \textbf{feasible} & \textbf{absent} & \textbf{change} \\
\midrule
Base diffusion &
50 (10.0$\pm$2.8) &
13 (2.6$\pm$1.0) &
8 (1.6$\pm$1.0) \\
Fixed control &
117 (23.4$\pm$14.1) &
72 (14.4$\pm$9.1) &
48 (9.6$\pm$7.9) \\
Delta-Seeker &
124 (24.8$\pm$14.6) &
100 (20.0$\pm$12.4) &
60 (12.0$\pm$6.4) \\
\bottomrule
\end{tabular}
\end{table}

\vspace{4pt}
\noindent\textbf{Verified Converter Designs.}
Four representative corpus-absent converters are examined in
Fig.~\ref{fig:discovered_converters} and
Table~\ref{tab:conv_parent_child}. All four are buck-family variants, consistent with the available device set and 2:1 step-down target.

\begin{itemize}[leftmargin=*, topsep=2pt, partopsep=0pt, itemsep=4pt, parsep=0pt]

\item \textbf{\emph{D1, parallel inductor.}}
The generated $L_1$ is connected in parallel with $L_0$. Its nearest training-corpus parent already satisfies all four corners, while the generated design raises median efficiency from 99.37\% to 99.60\%. Replacing $L_0 \parallel L_1$ with a single inductor having half the inductance and winding resistance reproduces the generated behavior, showing that the improvement comes from reduced winding loss rather than a new current path.

\item \textbf{\emph{D2, additional switching current path.}}
The generated capacitor $C_1$ carries substantial current during
the switching interval in which the nearest training-corpus parent
lacks a corresponding path. Waveform analysis therefore shows that the added connection changes the parent from a non-functioning topology into a converter satisfying all four operating corners. Unlike D1, the generated element changes the circuit conduction path rather than only its effective component value.

\item \textbf{\emph{D3 and D4, active but reducible modifications.}}
In D3, the generated elements establish a conduction path for $L_2$ in both switching phases and suppress the large voltage excursions of its parent, while D4 repairs a parent that produces no output. Equivalent-circuit reduction, however, maps both designs to a known buck with three parallel inductors. Their benefit comes from repairing defective parent structures rather than introducing a distinct converter function.

\end{itemize}

These examples show that corpus absence alone does not establish a new circuit function. D1 improves an existing mechanism, D3 and D4 repair defective parents but reduce to a known buck structure, while D2 introduces an additional switching current path that changes the operation of its nearest parent. This motivates the post-search circuit analysis used to distinguish graph novelty from changes in circuit behavior.

\section{Conclusion}
\label{sec:conclusion}

We presented Delta-Seeker, a closed-loop generative search framework for circuit design discovery in which outcomes from earlier candidates guide both evaluation effort and subsequent topology generation. In op-amp discovery, cross-generation memory and state-dependent control increase the yield of designs that survive independent re-sizing and supply-temperature verification, while controlled ablations show that feedback changes the generated topologies rather than only their downstream selection. In DC-DC converter discovery, nearest-parent simulation, waveform analysis, and equivalent-circuit reduction show why structural novelty alone is insufficient: generated modifications can improve an existing mechanism, repair a defective topology, or introduce a new conduction path that changes circuit operation. These results establish closed-loop feedback and post-search circuit verification as complementary elements for moving generative EDA beyond candidate generation toward reproducible circuit discovery.

\bibliographystyle{IEEEtran}
\bibliography{references}

\end{document}